# Enhancing Clinical Decision Support and Differential Diagnosis with Knowledge Graphs, and Retrieval Augmented Generation in Generative AI

**Henri Feto[1], Abicumaran Uthamacumaran[2], Hector Zenil[1,2]***

[1]Department of Biomedical Engineering, King's College London, University of London, London, United Kingdom

[2]Oxford Immune Algorithmics, Reading, United Kingdom

**Corresponding author: Hector Zenil*

## Abstract

Diagnostic error carries a substantial clinical and economic burden, while unconstrained large language models (LLMs) remain vulnerable to hallucination and weak integration of quantitative laboratory dynamics. We developed an integrated clinical decision-support pipeline combining disease-specific biomarker correlation graphs, coupled ordinary differential equations (ODEs), deep sequence classification and retrieval-augmented generation (RAG). For 103 hematological and systemic disease classes derived from a full blood count (FBC) repository, per-disease biomarker networks were used as coupling matrices to generate 30 stochastic trajectories per disease (3,090 trajectories total). A one-dimensional convolutional neural network (CNN) and long short-term memory (LSTM) network classified disease trajectories and six dynamical clusters. A constrained GPT-4o-mini RAG layer used a 19-pattern BMJ Best Practice/NICE corpus to generate four-section differential diagnoses, evaluated for diagnostic suitability, evidential grounding and clinical plausibility. Across five archived random-seed runs, disease-level accuracy was 0.940 ± 0.006 for the CNN (95% CI 0.933–0.948) and 0.852 ± 0.019 for the LSTM (95% CI 0.828–0.875); the paired CNN advantage was 8.87 percentage points (95% CI 6.47–11.27; paired $t(4)=10.26$, $p=5.1\times10^{-4}$; Hedges $g=3.67$). Among 100 sampled RAG cases, 96 parsed successfully; retrieved evidence was cited in 97.9% of parsed outputs, the true diagnosis was mentioned in 71.9%, and the multi-axis composite score was 3.82/5 with a 47.9% strict pass rate. The central empirical finding was a decoupling between grounding and diagnostic correctness: classifier-correct versus classifier-wrong outputs differed markedly in diagnostic suitability but not evidential grounding. Post-hoc analysis confirmed a 1.02-point diagnostic-score difference (Mann–Whitney $p=0.0024$; Hedges $g=0.72$), whereas grounding differed by only −0.02 points ($p=0.839$; $g=-0.04$). These results support structured quantitative representation before generation, while emphasizing that citation fidelity is not equivalent to diagnostic correctness and must be monitored separately in clinical RAG systems.

## Introduction

Differential diagnosis requires clinicians to distinguish diseases with overlapping presentations while integrating laboratory measurements, symptoms, imaging and history. Diagnostic error in UK practice is substantial: missed diagnostic opportunities account for a large fraction of malpractice claims, clinically important harm occurs in primary and emergency care, and NHS Resolution has reported billions of pounds in compensation and clinical-negligence liabilities [1–4]. The representation problem is especially apparent in laboratory medicine, where disease states are expressed through multivariate patterns rather than isolated analytes.

A standard FBC returns multiple interdependent hematological variables. Low haemoglobin, for example, has different diagnostic meaning depending on MCV and related red-cell indices, while a raised WBC count must be interpreted with the relative behaviour of neutrophils, lymphocytes and monocytes. Biomarkers are also dynamic: disease progression, treatment response and physiological compensation are trajectories rather than static snapshots [48]. Conventional protocol-based systems and many tabular machine-learning approaches do not explicitly preserve this combination of network dependence and temporal evolution [49].

LLMs such as GPT-4 and Med-PaLM have shown strong performance on standardized medical benchmarks, yet their performance is less reliable under realistic clinical conditions [5,6]. Hallucination and inflexible reasoning remain important failure modes [7,8]. In addition, numerical biomarker dynamics are typically converted into text before an LLM can reason over them, potentially discarding structure. These limitations motivate two complementary requirements: quantitative disease state should be represented in a structured dynamical form, and generative reasoning should be grounded in retrievable evidence.

Healthcare knowledge graphs offer interpretable representations of medical entities and their relations [9], while data-derived biomarker networks provide a complementary graph representation in which nodes are physiological variables and weighted edges encode statistical relationships [10]. RAG externalizes part of an LLM's knowledge base by retrieving relevant evidence before generation, improving factual traceability on knowledge-intensive tasks [11,12]. Network medicine and continuous-time disease modelling provide a further basis for coupling biomarker dependence to dynamical evolution [28–34].

The present work integrates these traditions. Disease-specific Pearson correlation networks are converted into coupled ODE systems that generate synthetic dynamical signatures. CNN and LSTM classifiers learn those signatures at fine-grained disease and coarse dynamical-cluster levels. At inference, trajectory summaries and classifier predictions are combined with semantic retrieval over a curated clinical corpus, after which a constrained LLM generates a source-cited differential diagnosis. The study then evaluates both classification performance and the distinct dimensions of generated-answer quality.

Our neurosymbolic framework combines structured and interpretable biomarker networks and dynamical rules with neural sequence learning and retrieval-grounded language reasoning. Conceptually, it progresses from a biomarker network to a dynamical system, then to a learned latent representation, and finally to RAG-based clinical reasoning, interlinking mechanistic structure discovery to data-driven prediction and evidence-grounded interpretation. The study also considered socioeconomic context, UK medical-device regulation, privacy, equity and diversity implications, and sustainability (See Suppl. Information). These materials are retained in Supplementary Note 1.

## Methods

### Study design, data and preprocessing

The source repository comprised 105 disease-labelled FBC profiles spanning hematological malignancies, anaemias, infections, electrolyte and endocrine disorders, and cardiovascular conditions. Each disease profile contained five patient samples and eleven CBC/FBC variables: WBC, lymphocytes, monocytes, segmented neutrophils, eosinophils, basophils, RBC, haemoglobin, MCV, platelet count and MPV. Two profiles were excluded because of missing rows, leaving 103 disease classes.

Preprocessing was performed globally and within disease. Columns with >50% missingness or cohort-wide standard deviation $<10^{-6}$ were removed; basophils were excluded because of negligible across-cohort variance, leaving ten biomarkers. Remaining global missing values were imputed with column means. Within each disease, residual missing values were replaced by zero and variables were z-score normalized. Zero-variance within-disease columns were perturbed with additive Gaussian noise of magnitude $10^{-6}$ before correlation estimation to avoid undefined correlations. The normalized input tensor therefore had dimensions 103×5×10.

### Disease-specific biomarker graphs and coupled ODE trajectories

For each disease d, the ten retained biomarkers were treated as graph nodes and the within-disease Pearson correlation matrix $R_d$ as a symmetric weighted adjacency representation. The diagonal was removed, $A_d=R_d-I$, and the matrix was spectrally rescaled as $W_d=(g/\rho(A_d))A_d$ with coupling scale g=0.95. The dynamical state x(t) followed the coupled nonlinear ODE $dx/dt=-\gamma x(t)+W_d \tanh(x(t))$, with $\gamma=0.5$ and initial state $x(0)\sim N(0,\sigma_0^2 I)$, $\sigma_0=0.1$. The system was integrated with LSODA (`scipy.integrate.odeint`) over T=10 arbitrary units at 100 time points with absolute and relative tolerances of $10^{-6}$. Thirty stochastic initial conditions were simulated per disease, yielding 3,090 trajectories of shape 100×10. Trajectories were clipped to [−5,5] as a numerical safeguard.

The coupling scale was selected to retain nontrivial but stable dynamics: substantially smaller coupling would drive rapid relaxation toward the fixed point, whereas coupling beyond the stability boundary could produce divergent or numerically unstable trajectories. The graph construction is data-derived and correlation based; it does not by itself establish causal edges.

### Dynamical clustering

Each 100×10 trajectory was flattened to a 1,000-dimensional feature vector and feature-wise standardized. K-means clustering was run with ten independent restarts. Elbow analysis over K=2–14 supported K=6 as a conservative choice near the point of diminishing inertia reduction. PCA projections were used qualitatively to inspect cluster structure. Cluster labels served as a coarse target alongside the 103-class disease labels.

### CNN and LSTM classifiers

The CNN treated biomarkers as channels and time as the spatial dimension. It used Conv1D layers with 32, 64 and 128 channels, kernel sizes 5, 5 and 3, ReLU activations, two max-pooling operations, adaptive global average pooling and a fully connected classification head with dropout 0.3. The LSTM consumed the 100×10 sequence directly through two recurrent layers with hidden dimension 64 and dropout 0.3, with classification from the final hidden representation. The archived disease checkpoints contain 49,927 trainable parameters for the CNN and 59,431 for the LSTM; the six-cluster checkpoints contain 37,414 and 53,126 parameters, respectively.

Both models were trained for 40 epochs using cross-entropy loss, Adam with initial learning rate 10⁻3 and cosine annealing. The checkpoint maximizing validation accuracy was retained. Data were stratified into 64% training, 16% validation and 20% test partitions, and per-biomarker normalization statistics were fit on the training partition only. Results were repeated across five random seeds.

CNN and LSTM were selected as complementary, sequence-learning algorithms with distinct inductive biases: the CNN captures local, translation/shift-invariant temporal motifs, whereas the LSTM models ordered, sequential long-range temporal dependencies. Their comparison therefore tests whether the ODE-derived disease signatures are better represented by local dynamical patterns or recurrent sequence memory without introducing the complexity of larger foundation architectures.

### Retrieval-augmented generation

The RAG stage consisted of corpus curation, semantic retrieval and constrained generation. A reference corpus of N=19 clinical biomarker-pattern entries was curated from BMJ Best Practice and NICE guidance. Each entry contained a disease-class label, a free-text description of a characteristic CBC/FBC signature and a citation string. Patterns were embedded with `all-MiniLM-L6-v2` into 384-dimensional sentence-transformer vectors [41].

At inference, each simulated trajectory was converted to a natural-language description by mapping start-to-end biomarker changes into categorical direction tokens. The query embedding was compared with corpus embeddings by cosine similarity and the top k=3 entries were retrieved. These entries and the classifier's top three predictions were inserted into a prompt instructing GPT-4o-mini to return four sections: most likely diagnosis, alternatives, supporting evidence and caveats. The prompt prohibited fabrication of sources and required citation only to retrieved patterns. Generation used temperature 0.2 and a 400-token output limit; the generated response itself was constrained to <250 words.

### Evaluation and statistical analysis

Classifier evaluation used held-out top-1 accuracy, macro/weighted F1 and confusion matrices at disease and cluster granularities. Generation quality was assessed through retrieval citation rate, true-diagnosis mention rate, LLM–classifier top-1 agreement and a multi-axis LLM-as-judge protocol [42]. GPT-4o scored each parsed differential on diagnostic suitability, evidential grounding and clinical plausibility using a 1–5 rubric. The composite was the unweighted mean; a strict pass required scores ≥4 on all three axes. The judge received the generated differential, true diagnosis and retrieved evidence and returned structured JSON.

Accuracy means were supplemented with Student-t 95% confidence intervals across the five paired seed runs. CNN–LSTM differences used paired t-tests with paired standardized effect sizes (Hedges g_z); exact Wilcoxon signed-rank p-values are also provided in Supplementary Table S4. Proportions were supplemented with Wilson 95% confidence intervals. For the 96 LLM-judge cases, classifier-correct versus classifier-wrong score distributions were compared with two-sided Mann–Whitney tests and Hedges g.

**Table 1 | Core pipeline specification.**

| Component | Specification |
|---|---|
| Data | 105 disease profiles; 2 excluded; 103 classes; 5 samples/disease; 10 retained biomarkers |
| Graph | Within-disease Pearson correlation network |
| ODE | γ=0.5; g=0.95; T=10; 100 steps; 30 initial conditions/disease; LSODA |
| CNN | Conv1D 32/64/128; kernels 5/5/3; max-pool; global average pooling; dropout 0.3 |
| LSTM | 2 layers; hidden dimension 64; dropout 0.3 |
| Training | 40 epochs; Adam $10^{-3}$; cosine annealing; 64/16/20 stratified split; five random seeds |
| RAG | all-MiniLM-L6-v2; N=19 corpus; top-k=3; GPT-4o-mini; T=0.2; 400-token cap |
| Generation evaluation | Citation, truth mention, classifier agreement, 3-axis LLM judge, strict pass ≥4/5 on all axes |

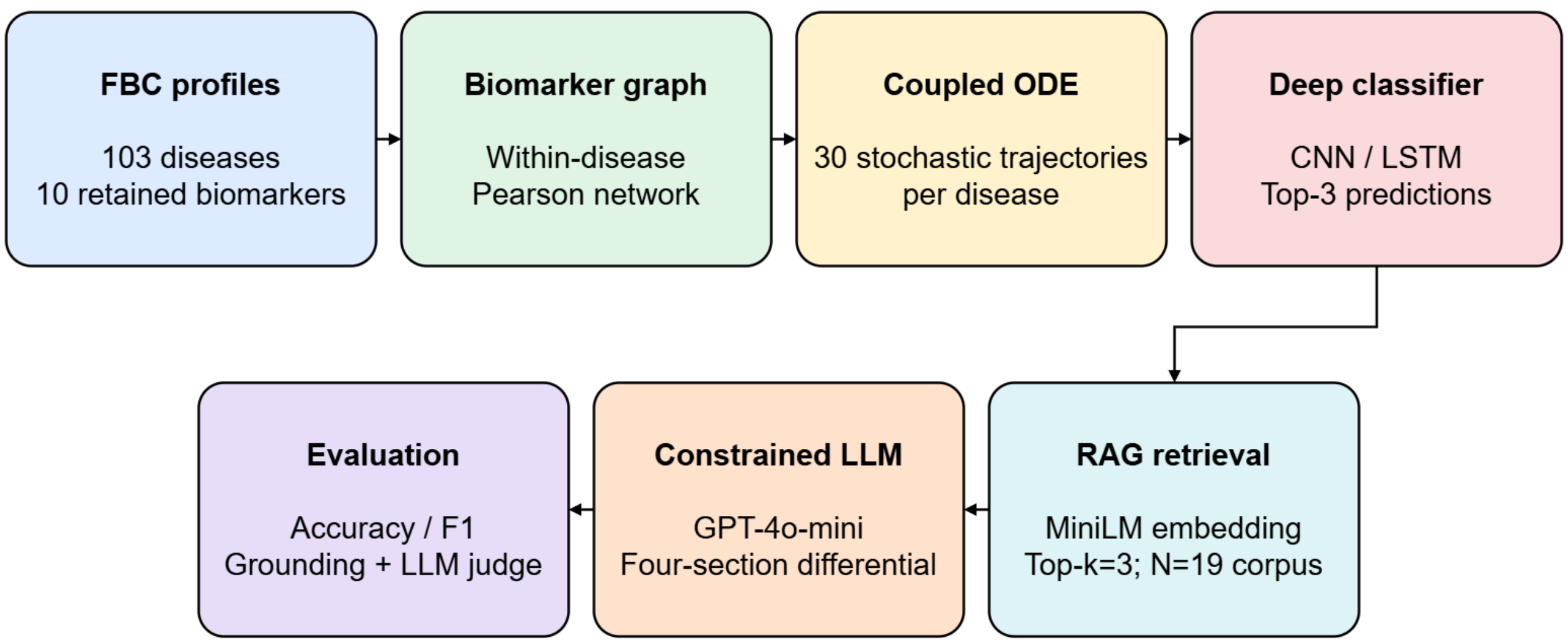

**Figure 1 | Neurosymbolic clinical decision-support workflow.** Disease-specific FBC profiles are converted to biomarker correlation graphs and coupled ODE trajectories, classified by deep sequence models, mapped to retrieved clinical patterns and then passed to a constrained LLM. Evaluation deliberately separates discriminative accuracy from evidential grounding and generated-answer quality.

## Results

### Disease-specific graph structure and dynamical archetypes

All 103 disease classes produced valid bounded trajectories under the spectrally rescaled ODE system; no disease was lost to integration failure. Per-disease correlation matrices and network views differed substantially. Microcytic anaemia showed prominent red-cell-line coupling, chronic lymphocytic leukaemia showed strong cross-line negative coupling involving WBC/lymphocytes and red-cell variables, and bacterial pneumonia showed a coupled inflammatory WBC–monocyte–neutrophil block. These examples illustrate that the disease representation is encoded in multivariate dependency structure rather than a single abnormal marker.

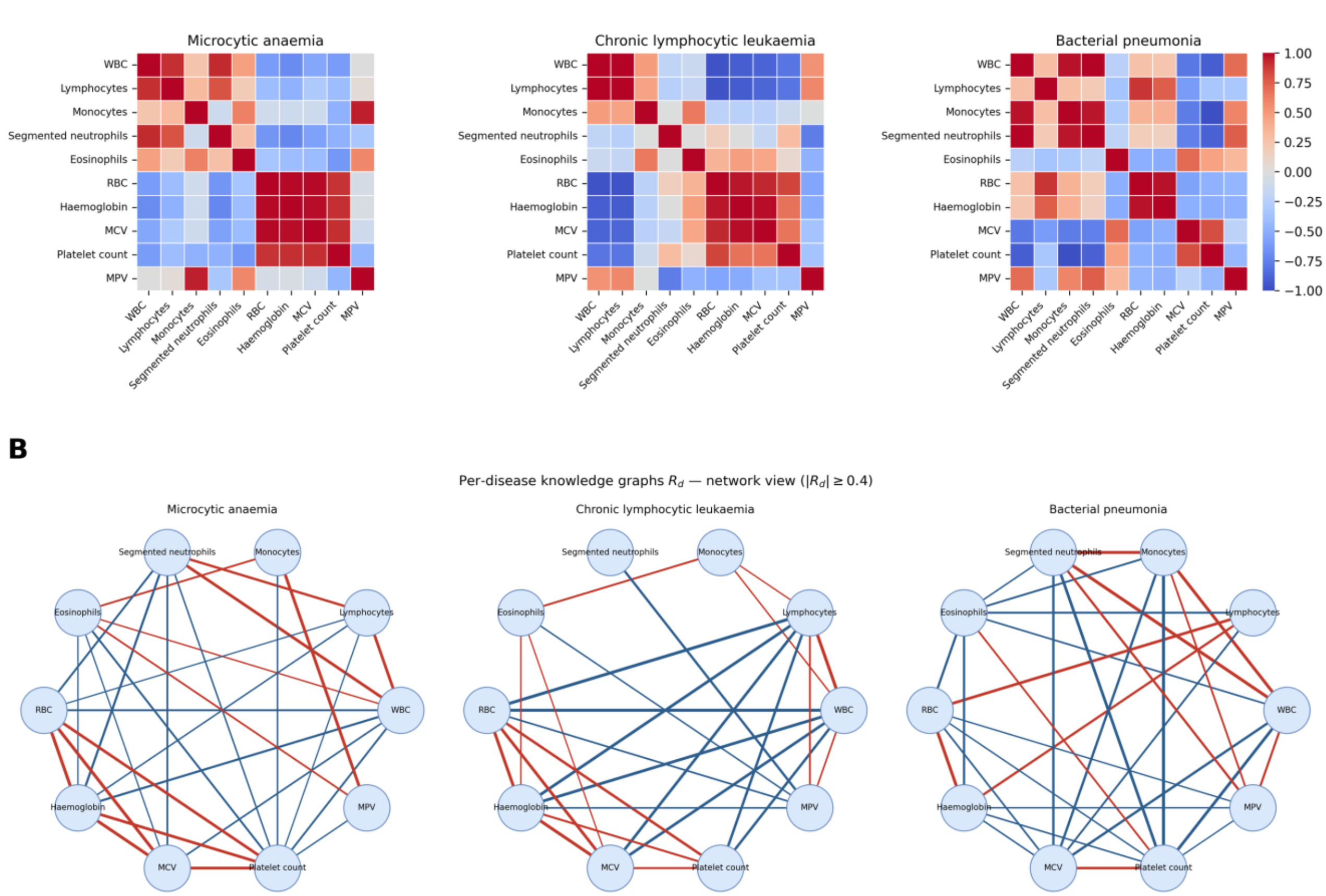

**Figure 2 | Exemplar disease-specific biomarker structure.** (A) Pearson correlation matrices for microcytic anaemia, chronic lymphocytic leukaemia and bacterial pneumonia. (B) Corresponding thresholded network views (|r|≥0.4 for visualization). Full network thumbnails for all 103 diseases are retained in Suppl. Info. *Figure S1).

The simulated trajectories were bounded and smoothly varying, with qualitatively distinct regimes including positive divergence, negative decline and near-zero saturation. K-means analysis of all 3,090 flattened trajectories showed diminishing reductions in inertia after approximately K=6; the silhouette score at K=6 was 0.122, indicating modest but non-random and overlapping structure. The six clusters were dynamical rather than taxonomic: clinical categories appeared in multiple clusters, no disease's 30 replicates fell entirely into one cluster, and the mean/median modal cluster purities were 44.6%/43.3%. Cluster 4 was the largest, containing 1,158 trajectories and serving as a broad modal/default dynamical group.

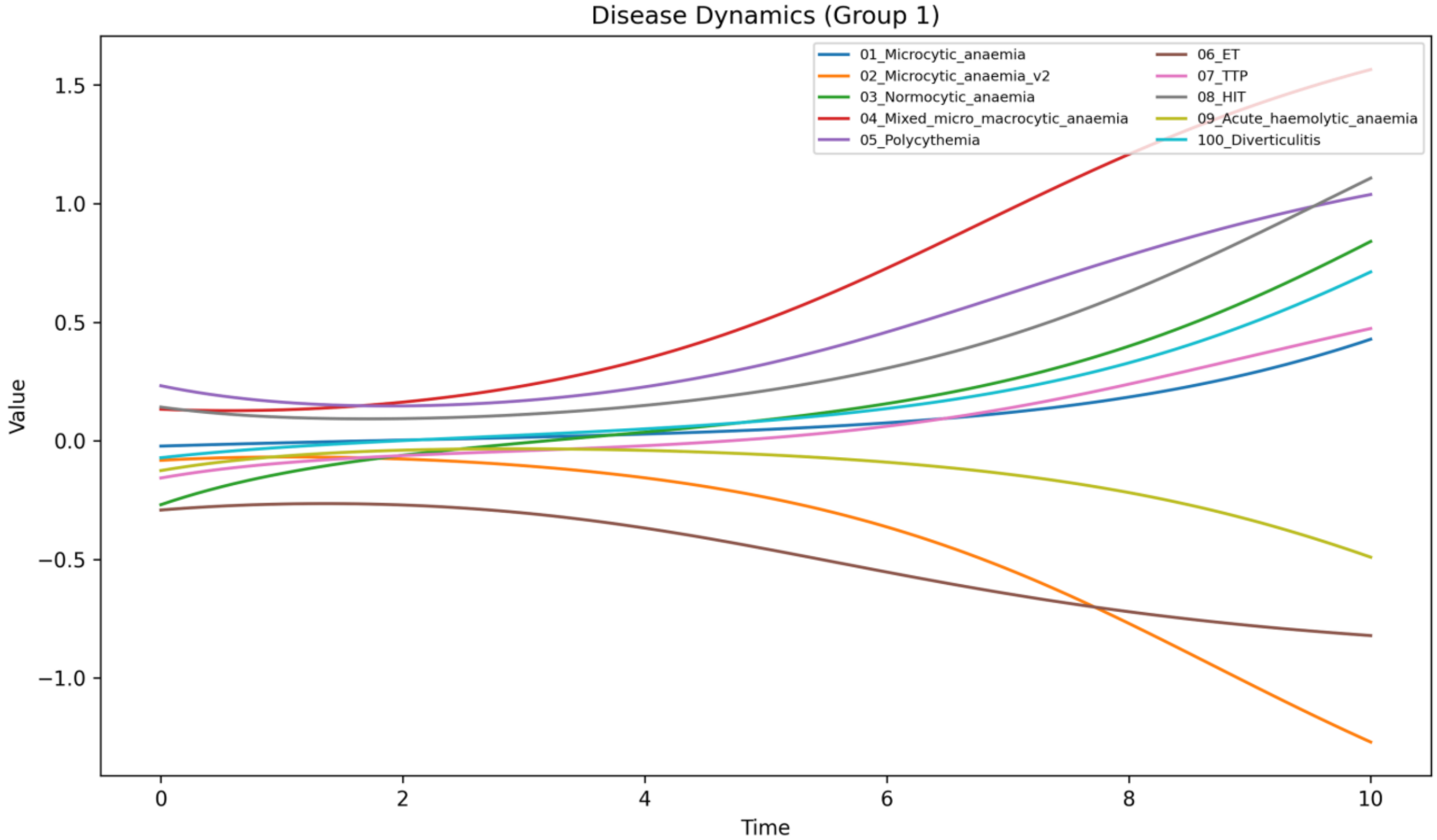

**B**

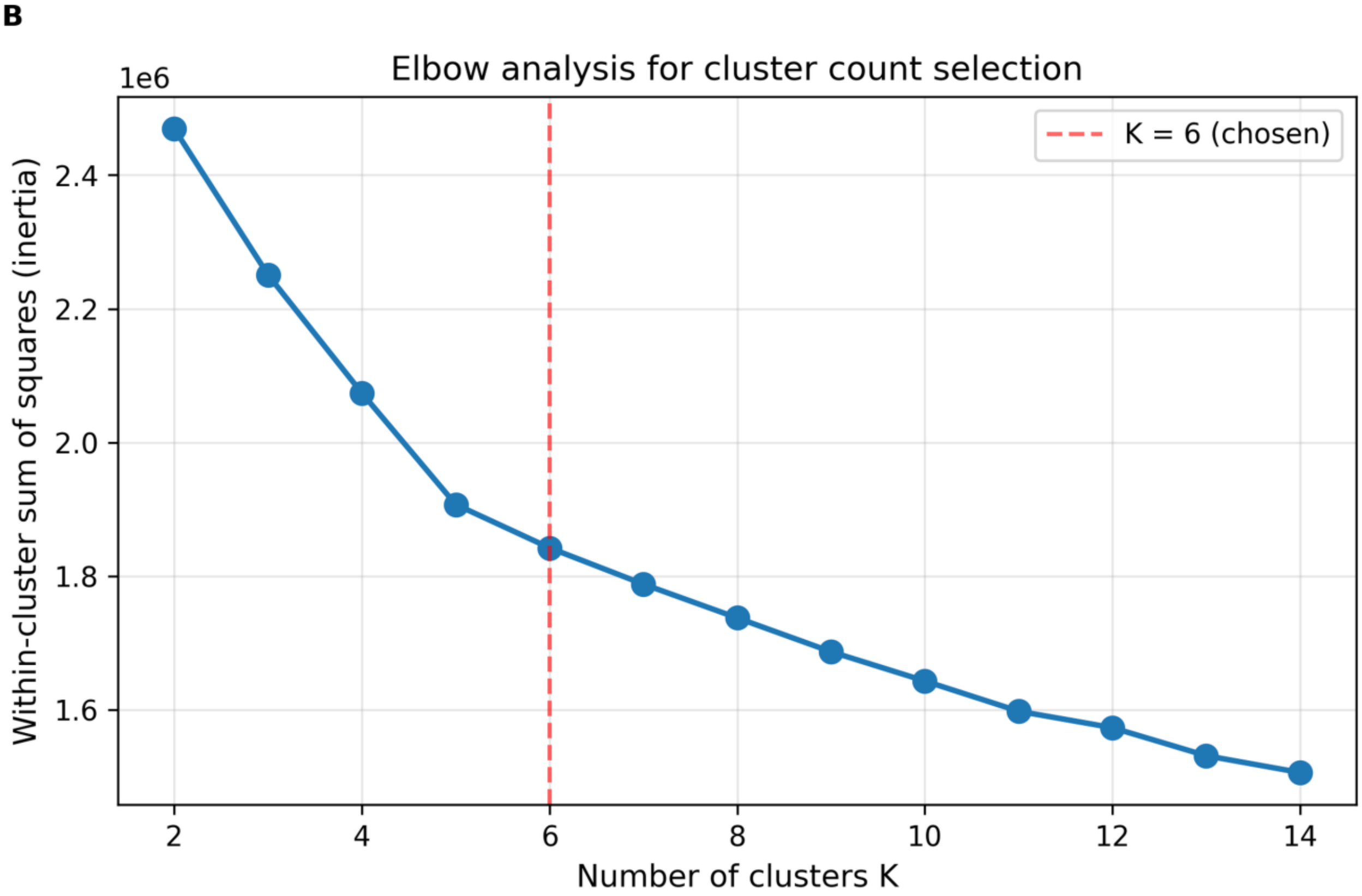

**Figure 3 | Simulated dynamical signatures and cluster selection.** (A) Representative disease trajectories illustrate divergent and convergent behaviours generated by the coupled ODE system. (B) Elbow analysis over K=2–14 supports six clusters as a conservative choice near the major inflection. All disease trajectory groups are provided in the Supplementary Info.

### CNN versus LSTM classification

Across the five random-seed runs, the CNN achieved mean disease-class accuracy of 0.940 compared with 0.852 for the LSTM. Seed-42 macro/weighted F1 values were 0.923/0.923 for the CNN and 0.803/0.803 for the LSTM. On the six-cluster task, accuracies were 0.939 for the CNN and 0.922 for the LSTM, with macro F1 values of 0.923 and 0.919, respectively. The narrower model gap at cluster level is consistent with the LSTM approaching CNN performance as label granularity becomes coarser.

The reanalysis quantified uncertainty around these means. CNN disease accuracy was 0.9405 ± 0.0060 (95% CI 0.9330–0.9479) versus 0.8518 ± 0.0190 (95% CI 0.8282–0.8753) for the LSTM. The paired mean difference was 0.0887 (95% CI 0.0647–0.1127), paired $t(4)=10.26$, $p=0.00051$, Hedges $g_z=3.67$. On the cluster task, the CNN–LSTM difference was 0.0168 (95% CI −0.0028–0.0364), $t(4)=2.38$, $p=0.0758$, Hedges $g_z=0.85$. Exact two-sided Wilcoxon $p=0.0625$ for both comparisons, reflecting the coarse p-value resolution with only five pairs.

The CNN confusion matrix was strongly diagonal. Residual errors concentrated among disease variants with similar FBC signatures, including thrombocytopenia and hyponatraemia subtypes. The LSTM showed more off-diagonal mass and class-collapse-like vertical streaks. At cluster level, both matrices were strongly diagonal, but the CNN concentrated residual errors in the dominant cluster 4 whereas the LSTM distributed errors more diffusely.

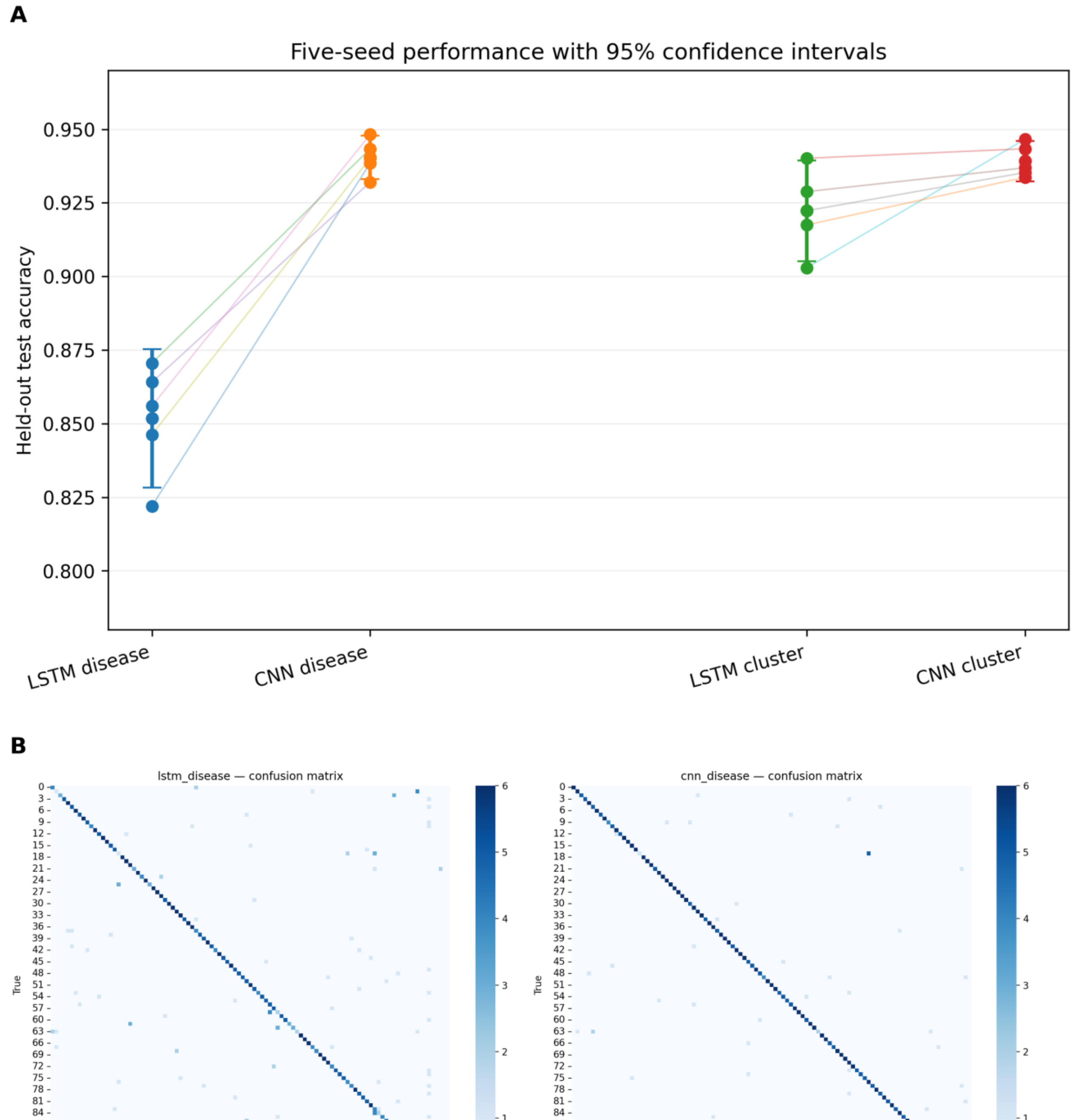

**Figure 4 | Classifier performance.** (A) Five paired random-seed test accuracies with mean 95% confidence intervals. Lines join matched seed runs. (B) Disease-level confusion matrices for the LSTM and CNN show substantially stronger diagonal concentration for the CNN.

**Table 2 | Deep-classifier performance with post-hoc uncertainty estimates.**

| Model | Task | Accuracy mean±SD | 95% CI across seeds | Macro F1 (seed 42) |
|---|---|---|---|---|
| LSTM | Disease (103 classes) | 0.852±0.019 | 0.828–0.875 | 0.803 |
| CNN | Disease (103 classes) | 0.940±0.006 | 0.933–0.948 | 0.923 |
| LSTM | Cluster (6 classes) | 0.922±0.014 | 0.905–0.939 | 0.919 |
| CNN | Cluster (6 classes) | 0.939±0.006 | 0.932–0.946 | 0.923 |

## Retrieval-augmented differential diagnosis

The full RAG pipeline was evaluated on a stratified sample of 100 test trajectories. Ninety-six outputs parsed successfully under the structured response protocol; four failed JSON decoding and were excluded from judge analyses. Among the 96 parsed outputs, the retrieval citation rate was 97.9%, the true diagnosis was mentioned in 71.9%, and LLM–classifier top-1 agreement was 75%. The upstream classifier was correct in 91.7% of the same cases. Mean generated length was 219 words, within the 250-word prompt limit.

Wilson 95% confidence intervals were 90.2–98.4% for JSON parse success (96/100), 92.7–99.4% for retrieval citation (94/96), 62.2–79.9% for true-diagnosis mention (69/96), 65.5–82.6% for LLM–classifier agreement (72/96), and 84.4–95.7% for upstream classifier correctness (88/96). The lower true-diagnosis mention rate relative to upstream classifier accuracy indicates that the generator does not simply repeat classifier predictions. When retrieved textual patterns support an alternative, the LLM can promote that alternative above a correct classifier output. This retrieval-over-classifier override was the principal observed failure mode and is illustrated by the polymyalgia rheumatica case retained in Supplementary Note 5.

## Multi-axis LLM evaluation and classifier–grounding decoupling

Across 96 parsed outputs, mean judge scores were 4.06 for diagnostic suitability, 3.60 for evidential grounding and 3.80 for clinical plausibility; the composite mean was 3.82/5. Forty-six outputs (47.9%) passed the strict criterion of ≥4 on all three axes. Score distributions differed by dimension: diagnostic suitability was polarized toward very high or low values, grounding was concentrated around 3–4, and plausibility was intermediate.

The corresponding mean±SD and 95% CIs were 4.06 ± 1.43 (3.77–4.35) for diagnostic suitability, 3.60±0.57 (3.49–3.72) for evidential grounding, 3.80±0.87 (3.63–3.98) for clinical plausibility, and 3.82±0.85 (3.65–4.00) for the composite. The strict-pass Wilson 95% CI was 38.2–57.8%. The strongest behavioural result emerged after stratifying by whether the upstream classifier was correct. On 88 classifier-correct cases, diagnostic suitability averaged 4.15 and plausibility 3.85; on eight classifier-wrong cases these means fell to 3.13 and 3.25. Evidential grounding, in contrast, was 3.60 versus 3.63. Thus, the generator remained similarly faithful to what it retrieved even when the overall diagnostic pathway was wrong. Post-hoc inferential analysis supports this dissociation. Diagnostic suitability differed by 1.02 points (95% CI 0.17–1.88; Mann–Whitney $p=0.0024$; Hedges $g=0.72$) and plausibility by 0.60 points (95% CI 0.19–1.01; $p=0.019$; $g=0.70$). Grounding differed by only −0.023 points (95% CI −0.46–0.42; $p=0.839$; $g=-0.04$). The composite difference was 0.53 points (95% CI 0.14–0.92; $p=0.019$; $g=0.63$).

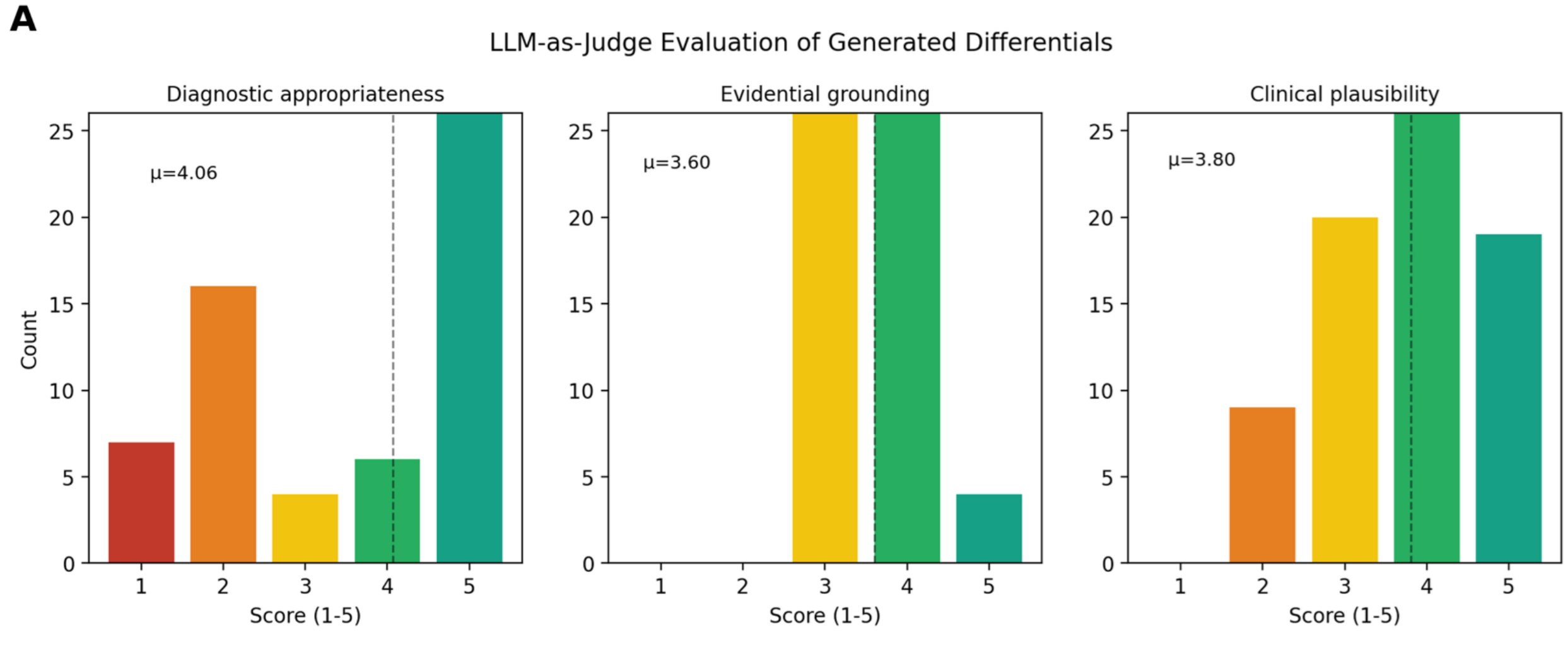

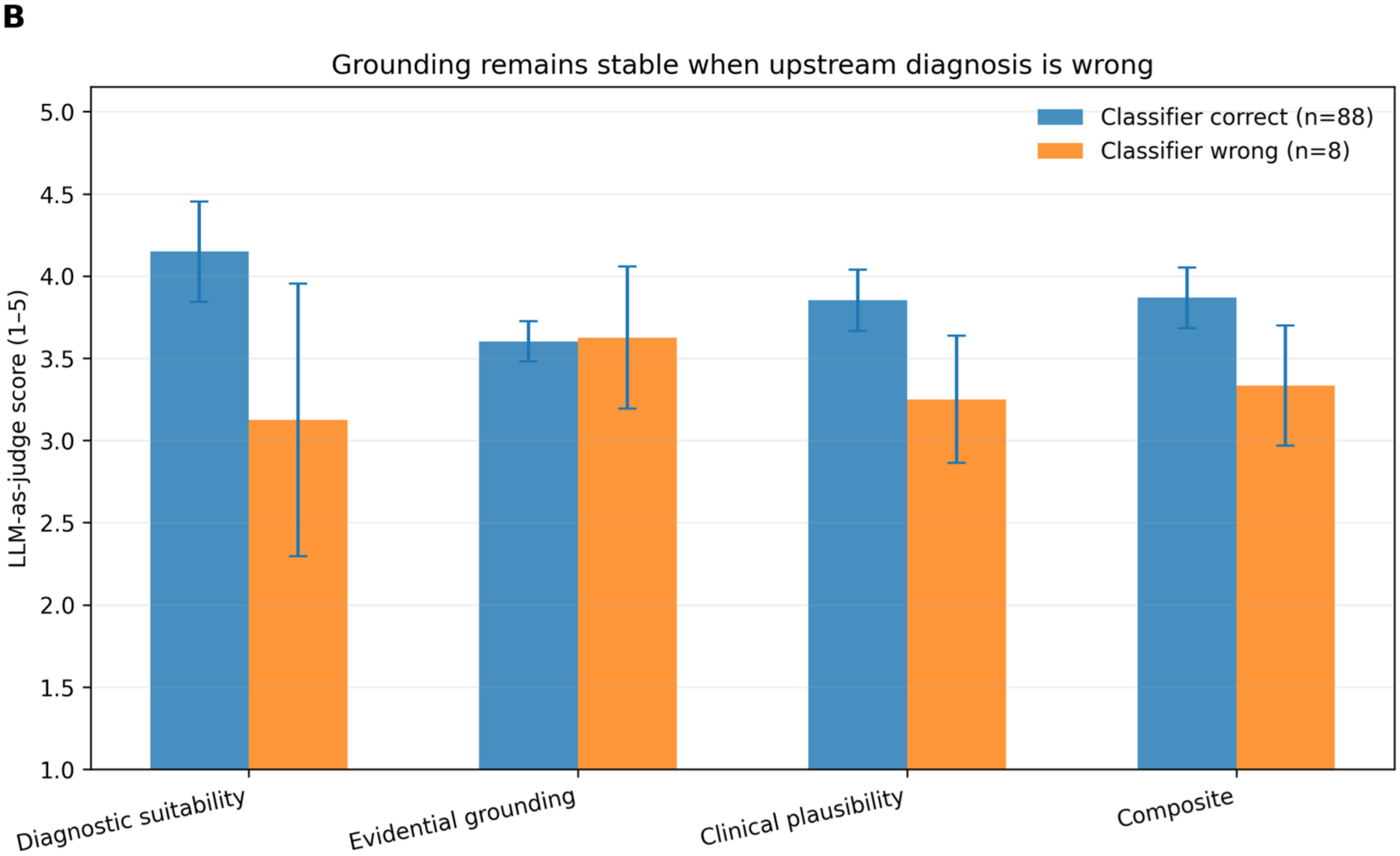

**Figure 5 | LLM-as-judge evaluation.** (A) Score distributions for diagnostic suitability, evidential grounding and clinical plausibility. (B) New post-hoc stratification by upstream classifier correctness with 95% confidence intervals. Diagnostic and plausibility scores fall when the classifier is wrong, whereas grounding is nearly unchanged.

## Table 3 | RAG and LLM evaluation.

| Metric | Estimate | 95% CI / comparison |
|---|---|---|
| JSON parse success | 96/100 (96.0%) | 90.2–98.4% |
| Retrieval citation | 94/96 (97.9%) | 92.7–99.4% |
| True diagnosis mentioned | 69/96 (71.9%) | 62.2–79.9% |
| LLM–classifier top-1 agreement | 72/96 (75.0%) | 65.5–82.6% |
| Strict LLM-judge pass | 46/96 (47.9%) | 38.2–57.8% |
| Diagnostic score: classifier correct vs wrong | 4.15 vs 3.13 | $\Delta$=1.02; p=0.0024; g=0.72 |
| Grounding score: classifier correct vs wrong | 3.60 vs 3.63 | $\Delta$=−0.02; p=0.839; g=−0.04 |
| Plausibility score: classifier correct vs wrong | 3.85 vs 3.25 | $\Delta$=0.60; p=0.019; g=0.70 |

## Discussion

This work developed a clinical decision-support architecture in which structured quantitative representation precedes generative reasoning. The per-disease FBC correlation graph defines a coupled dynamical system, synthetic trajectories expose disease-specific signatures to deep classifiers, and a retrieval-constrained LLM translates these quantitative predictions into an evidence-linked differential diagnosis. The approach addresses two distinct weaknesses of stand-alone LLMs: numerical biomarker structure is handled outside the language model, while generated claims are anchored to an external corpus.

The disease-classification result is notable because 103 labels were separated with a mean CNN accuracy near 0.94 using only 30 simulated trajectories per disease. Similar CNN performance on the six-cluster task suggests that disease-specific information remains strong despite the stochastic ODE initialization, while the soft K-means structure indicates that the clusters are not substitutes for disease taxonomy. The CNN's advantage over the LSTM is consistent with local temporal motifs and translation-tolerant pattern extraction being well matched to the simulated trajectories; the LSTM converged more slowly and showed greater seed-to-seed variability.

The generative layer adds a different type of value. A 97.9% citation rate shows that prompt-level grounding constraints can be followed consistently, yet the true diagnosis was mentioned less often than the classifier was correct. This gap is not merely an error rate; it demonstrates that a RAG model can be evidentially faithful to an incorrect retrieval path. The classifier-grounding decoupling therefore has practical importance: citation fidelity, diagnostic correctness and clinical plausibility should be monitored as separate deployment metrics rather than collapsed into a single quality score.

Further, the grounding difference between classifier-correct and classifier-wrong cases is essentially null ($g \approx -0.04$; $p=0.839$), while diagnostic suitability and plausibility show moderate-to-large, standardized differences. This supports the study's claim as an empirically measurable property of the current constrained RAG pipeline rather than an impression drawn only from example cases.

Several limitations constrain clinical interpretation. First, each disease graph is estimated from only five patient profiles, limiting the reliability of correlation estimates and precluding robust partial-correlation or graphical-lasso estimation. Second, the ODE trajectories are synthetic representations of covariance structure rather than observed patient time courses; their directional polarity does not always reproduce literal clinical progression. This creates an asymmetry in which the classifier can learn the synthetic signature while the LLM interprets a textualized direction summary against clinical descriptions. Third, the 19-pattern retrieval corpus covers only a subset of the 103 disease classes, increasing the risk that semantically plausible but diagnostically incorrect patterns dominate generation. Fourth, all multi-axis scores come from a single LLM judge, without clinician adjudication or inter-rater reliability.

Additional methodological limitations are that other architecture baselines were not evaluated. Future studies should evaluate our findings against foundation and world models, and other types of recurrent neural networks (such as reservoir computing), joint-embedding architectures and transformers. These analyses should be performed before a high-stakes clinical-AI submission with the raw trajectory tensor and latent outputs can be re-exported.

The Supplementary Information additionally discusses UK SaMD regulation, AI Airlock, UK GDPR, equity concerns in reference ranges and medical LLMs, NHS cost context, and sustainability. These considerations remain relevant because the present system is a research prototype, not a deployable medical device. Real-world translation would require population-specific validation, longitudinal EHR data, larger disease cohorts, broader retrieval coverage, clinician-in-the-loop evaluation, and regulatory and equity audits. These materials are retained in the Supplementary Information [13–18,38–40,43].

Future work follows directly from these limitations. Larger disease cohorts would permit sparse/partial-correlation graphs; real longitudinal FBC series would test whether the ODE representation captures observed progression and enable causal-discovery extensions; and multi-judge plus clinician evaluation would permit formal inter-rater statistics such as Cohen's kappa $\kappa$ or Krippendorff's alpha $\alpha$. A Transformer comparator and embedding-alignment analysis would test whether the observed CNN advantage is architecture-specific and whether graph geometry, classifier latent space and RAG retrieval space encode concordant disease separability.

More broadly, the present framework points toward a convergence between dynamical systems modelling and machine learning, in which learned representations are evaluated not only by predictive performance but also by the causal structure of the dynamics (i.e., information geoemtries) that they encode. The coupled-ODE construction used here provides an explicit dynamical substrate, whereas the CNN and LSTM learn latent representations of the resulting trajectories. Future studies should therefore compare these representations across broader AI architectures, including Transformers, RNNs, neural ODEs, graph-based neural networks, state-space and joint-embedding/world models, and determine whether their latent spaces preserve dynamical properties such as attractor organization, transition structure, stability, bifurcations at critical thresholds, and perturbation sensitivity. Such comparisons could help distinguish architectures that merely classify simulated trajectories from those that recover compact representations of the underlying generative dynamics.

A powerful direction to achieve this is to analyse these learned representations through Algorithmic Information Dynamics (AID). AID combines algorithmic information theory with systematic perturbation analysis to estimate the contribution of individual components to the shortest generative description of a complex system and has been proposed as a framework for causal discovery, causal deconvolution and dynamical reprogramming [44–47]. Rather than treating correlation, attention or latent-space proximity as sufficient evidence of inferred mechanism, an AID analysis could perturb biomarkers, graph edges, trajectory dimensions or latent features and quantify the resulting change in estimated algorithmic complexity (minimal description length; MDL). From AID's perspective, variables whose perturbation substantially changes the MDL of the system may carry greater generative or causal information than variables that are merely statistically associated. Applied across the graph network, dynamical and learned latent representations developed here, this could test whether CNN, LSTM, or other architectures to be evaluated in the future such as Transformer attention layers or world-model embeddings recover the same candidate causal structure as the biomarker network, identify algorithmically informative disease-state transitions. Further we can assess whether the algorithmic learning progressively compresses trajectories toward simpler generative descriptions while retaining disease-discriminative structure.

## Conclusion

This study integrates ML-augmented disease-specific biomarker networks, nonlinear dynamical simulation, deep classification and retrieval-grounded generative reasoning into a single clinical decision-support framework. Across 103 disease classes, the CNN achieved approximately 0.94 held-out accuracy across five random-seed runs, while the RAG layer produced source-cited differentials with high citation adherence. The central contribution is the quantitative separation of evidential grounding from diagnostic correctness: the generator can remain faithful to retrieved evidence even when the upstream diagnostic path is wrong. Clinical RAG evaluation should therefore treat grounding, correctness and plausibility as distinct operational properties. Larger patient cohorts, real longitudinal validation, formal cross-validation/benchmark expansion, broader corpus coverage and clinician adjudication are required before clinical deployment.